\documentclass[sigplan,10pt,prologue,dvipsnames,svgnames,shortlabels]{acmart}
\acmConference[PL'18]{ACM SIGPLAN Conference on Programming Languages}{January 01--03, 2018}{New York, NY, USA}
\acmYear{2018}
\acmISBN{} 
\acmDOI{} 
\startPage{1}

\setcopyright{none}

\usepackage{booktabs}   
\usepackage{subcaption} 

\usepackage{hyperref}
\hypersetup{
    colorlinks=true,
    linkcolor=blue,
    filecolor=magenta,      
    urlcolor=blue,
}

\usepackage{graphicx}
\graphicspath{{pics/}}

\usepackage{enumitem}
\usepackage{comment}
\usepackage{listings}
\usepackage{enumitem}

\begin{document}

\title{Unexpected Scaling in Path Copying Trees}


\author{Ilya Kokorin}
\affiliation{
  \institution{ITMO University}            
  \country{Russia}                    
}
\email{kokorin.ilya.1998@gmail.com}          

\author{Alexander Fedorov}

\affiliation{
   \institution{IST Austria}
   \country{Austria}
}
\email{afedorov2602@gmail.com} 

\author{Trevor Brown}
\affiliation{
  \institution{University of Waterloo, Canada}            
  \country{Canada}                    
}
\email{trevor.brown@uwaterloo.ca} 

\author{Vitaly Aksenov}
\affiliation{
  \institution{ITMO University, Russia}            
  \country{Russia}                    
}
\email{aksenov.vitaly@gmail.com}          

\begin{abstract}
Although a wide variety of handcrafted concurrent data structures have been proposed, there is considerable interest in universal approaches (henceforth called Universal Constructions or UCs) for building concurrent data structures. These approaches (semi-)automatically convert a sequential data structure into a concurrent one. The simplest approach uses locks that protect a sequential data structure and allow only one process to access it at a time. The resulting data structures use locks, and hence are blocking. Most work on UCs instead focuses on obtaining non-blocking progress guarantees such as obstruction-freedom, lock-freedom, or wait-freedom. Many non-blocking UCs have appeared. Key examples include the seminal wait-free UC by Herlihy, a NUMA-aware UC by Yi et al., and an efficient UC for large objects by Fatourou et al.

We borrow ideas from persistent data structures and multi-version concurrency control (MVCC), most notably path copying, and use them to implement concurrent versions of sequential persistent data structures. Despite our expectation that our data structures would not scale under write-heavy workloads, they scale in practice. We confirm this scaling analytically in our model with private per-process caches.
\end{abstract}



\maketitle

\section{Introduction}


Although a wide variety of handcrafted concurrent data structures have been proposed, there is considerable interest in universal approaches (henceforth called \emph{Universal Constructions} or UCs) for building concurrent data structures. These approaches (semi-)automatically convert a sequential data structure into a concurrent one. 
The simplest approach uses locks~\cite{herlihy2020art,lamport1987fast} that protect a sequential data structure and allow only one process to access it at a time.
The resulting data structures use locks, and hence are blocking.
Most work on UCs instead focuses on obtaining non-blocking progress guarantees such as 
\emph{obstruction-freedom}, \emph{lock-freedom} or \emph{wait-freedom}.
Many non-blocking UCs have appeared.
Key examples include 
the seminal wait-free UC~\cite{herlihy1991wait} by Herlihy, a NUMA-aware UC~\cite{yi2021universal} by Yi et~al., and an efficient UC for large objects~\cite{fatourou2020efficient} by Fatourou et~al.

In this work, we consider the simpler problem of implementing \textit{persistent} (also called \textit{functional}) data structures, which preserve the old version
whenever the data structure is modified~\cite{okasaki1999purely}.
Usually this entails copying a part of the data structure, for example, the path from the root to a modified node in a tree~\cite{kaplan2018persistent}, so that none of the existing nodes need to be changed directly.

We borrow ideas from persistent data structures and multi version concurrency control (MVCC)~\cite{sun2019supporting}, most notably path copying, and use them
to implement concurrent versions of sequential persistent data structures. 
Data structures implemented this way can be highly efficient for searches, but we expect them to not scale in write-heavy workloads.
Surprisingly, we found that a concurrent treap implemented in this way obtained up to 2.4x speedup compared to a sequential treap~\cite{seidel1996randomized} with 4 processes in a write-heavy workload. 
We present this effect experimentally, and analyze it in a model with private per-processor caches: informally, as the number of processes grows large, speedup in our treap of size $N$ tends to $\Omega(\log N)$.


\section{Straightforward Synchronization for Persistent Data Structures}
\label{construction-description-section}




In the following discussion, we focus on \textit{rooted} data structures, but one could imagine generalizing these ideas by adding a level of indirection in data structures with more than one \textit{entry point} (e.g., one could add a dummy root node containing all entry points).

We store a pointer to the current version of the persistent data structure (e.g., to the root of the current version of a persistent tree) in a \texttt{Read/CAS} register called \texttt{Root\_Ptr}.

%

Read-only operations (queries) read the current version and then execute sequentially on the obtained version.
Note that no other process can modify this version, so the sequential operation is trivially atomic. 

Modifying operations are implemented in the following way: 1)~read the current version; 2)~obtain the new version by applying the sequential modification using path copying (i.e., by copying the root, and copying each visited node); 3)~try to atomically replace the current version with the new one using \texttt{CAS}; if the \texttt{CAS} succeeds, return: the modifying operation has been successfully applied; otherwise, the data structure has been modified by some concurrent process: retry the execution from step (1).
This approach clearly produces a lock-free linearizable data structure.



We expect read-only operations to scale extremely well. Indeed, two processes may concurrently read the current version of the persistent data structure
and execute read-only persistent operations in parallel.

However, modification operations seemingly afford no opportunity for scaling.
When multiple modifications contend, only one can finish successfully, and the others must retry.
For example, consider concurrent modification operations on a set: 1)~process \texttt{P} calls \texttt{insert(2)} and fetches the current pointer \texttt{RP}; 2)~process \texttt{Q} calls \texttt{remove(5)} and fetches the current pointer \texttt{RP}; 3)~\texttt{P} constructs a new version $\texttt{RP}_{\texttt{P}}$ with key \texttt{2}; 4)~\texttt{Q} constructs a new version $\texttt{RP}_{\texttt{Q}}$ without key \texttt{5}; 5)~\texttt{P} successfully executes \texttt{CAS(\&Set.Root\_Pointer, RP, $\texttt{RP}_{\texttt{P}}$)}; 6)~\texttt{Q} executes \texttt{CAS} from \texttt{RP} to $\texttt{RP}_{\texttt{Q}}$ but fails; thus, \texttt{Q} must retry its operation.

    
    
    
    
    

Successful modifications are applied sequentially, one after another.
Intuitively, this should not scale at all in a workload where all operations must perform successful modifications. 
As we will see in Section~\ref{experiments-section}, this intuition would be incorrect.

\section{Analysis}
\label{analysis-section}

\begin{figure}
  \centering
  \caption{The new version (green) of the tree shares its nodes with the old version (white)}
  \includegraphics[width=0.7\linewidth]{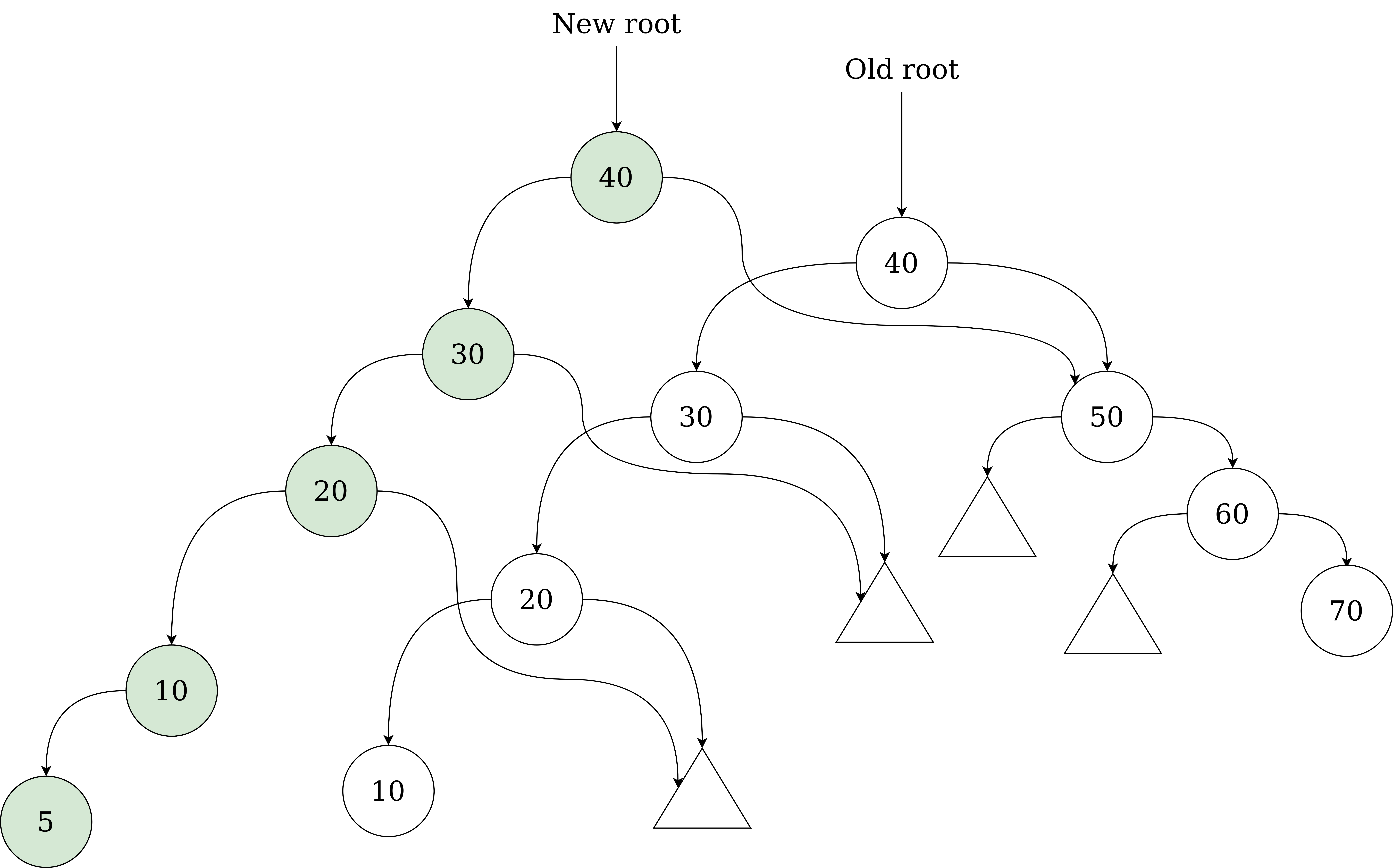}
  \label{tree-new}
\end{figure}

The key insight is that failed attempts to perform updates load data into processor caches that may be useful on future attempts.
To better understand, consider the binary search tree modification depicted in Fig.~\ref{tree-new}.
Suppose we want to insert two keys: \texttt{5} and \texttt{75}.
We compare how these insertions are performed sequentially and concurrently.


At first, we consider the sequential execution.
We insert key \texttt{5} into the tree.
It should be inserted as a left child of \texttt{10}.
Thus, we traverse the tree from the root to the leaf \texttt{10}.
On the way, we fetch nodes \texttt{\{40, 30, 20, 10\}} into the processor's cache.
Note this operation performs four uncached loads. 

Now, we insert \texttt{75}. 
It should be inserted as the right child of \texttt{70}.
Our traversal loads four nodes: \texttt{\{40, 50, 60, 70\}}.
Node \texttt{40} is already cached, while three other nodes must be loaded from memory.
Thus, we perform three uncached loads, for a total of seven uncached loads.

Now, we consider a concurrent execution with two processes, in which \texttt{P} inserts \texttt{5} and \texttt{Q} inserts \texttt{75}.
Initially, both processes read \texttt{Root\_Ptr} to load the current version.
Then, 1)~\texttt{P} traverses from the root to \texttt{10}, loading nodes \texttt{\{40, 30, 20, 10\}}, and 2)~\texttt{Q} traverses from the root to \texttt{70}, loading nodes \texttt{\{40, 50, 60, 70\}}.


Each process constructs a new version of the data structure, and tries to replace the root pointer using \texttt{CAS}. 
Suppose \texttt{P} succeeds and \texttt{Q} fails.
\texttt{Q} retries the operation, but on the \textit{new} version (Fig.~\ref{tree-new}).
Note that the new version shares most nodes with the old one.

\texttt{Q} inserts \texttt{75} into the new version.
Again, the key should be inserted as the right child of \texttt{70}.
\texttt{Q} loads four nodes \texttt{\{40, 50, 60, 70\}} from the new version of the tree.
Crucially, nodes \texttt{\{50, 60, 70\}} are already cached by \texttt{Q}.
This retry only incurs one cache miss!

Thus, there are only five serialized loads in the concurrent execution, compared to seven in the sequential execution.

\subsection{High-level analysis}
We use a simple model that allows us to analyze this effect.
(The full proof appears in Appendix~\ref{math-section}.)
In this model, the processes are synchronous, i.e., they perform one primitive operation per tick, and each process has its own cache of size $M$.
We show that for a large number of processes $P$, the speedup is $\Omega(\log N)$, where $N$ is the size of the tree.

Now, we give the intuition behind the proof.
To simplify it, we suppose that the tree is external and balanced, i.e., each operation passes though $\log N$ nodes.
We also assume that the workload consists of successful modification operations on keys chosen uniformly at random.
We first calculate the cost of an operation for one process: $(\log N - \log M) \cdot R + \log M$ where $M = O(N^{1 - \varepsilon})$ is the cache size and $R$ is the cost of an uncached load.
This expression captures the expected behaviour under least-recently-used caching.
The process should cache the first $\log M$ levels of the tree, and thus, $\log M$ nodes on a path are in the cache and $\log N - \log M$ are not.

To calculate the throughput in a system with $P$ processes, we suppose that $P$ is quite large ($\approx\Omega(min(R, \log N))$).
Thus, each operation performs several unsuccessful attempts, ending with one successful attempt, and all successful attempts (over all operations) are serialized.
Since the system is synchronous, each operation attempt $A$ loads the version of the data structure which is the result of a previous successful attempt $A'$.
The nodes evicted since the beginning of $A$ are those created by $A'$.
One can show that in expectation only two nodes on the path to the key are uncached.
Finally, the successful attempt of an operation incurs cost $2 \cdot R + (\log N - 2)$.
Since successful attempts are serialized, the expected total speedup is $\frac{(\log N - \log M) \cdot R + \log M}{2 \cdot R + (\log N - 2)}$ giving $\Omega(\log N)$ with $R = \Omega(\log N)$.

\section{Experiments}
\label{experiments-section}

We implemented a lock-free treap and ran experiments comparing it with a sequential treap in Java on a system with an 18 core Intel Xeon 5220.
Each data point is an average of 15 trials. 
We highlight the following two workloads.
(More results appear in Appendix~\ref{sec:aux:exp}.)

\subsection{Batch inserts and batch removes}

Suppose we have $P$ concurrent processes in the system. Initially the set consists of $10^6$ random integer keys. 
Processes operate on mutually disjoint sets of keys. 
Each process repeatedly: inserts all of its keys, one by one, then removes all of its keys. 
%
Since the key sets are disjoint, each operation successfully modifies the treap.
We report the \textit{speedup} for our treap over the sequential treap below.

\subsection{Random inserts and removes}

In this workload, we first insert $10^6$ random integers in \texttt{[$-10^6$; $10^6$]}, then each process repeatedly generates a random key and tries to insert/remove it with equal probability. 
Some operations do not modify the data structure (e.g., inserting a key that already exists).

\begin{center}
\begin{tabular}{|c|c|c|c|c|c|} 
 \hline
 Workload & Seq Treap & UC 1p & UC 4p & UC 10p & UC 17p \\
 \hline
 Batch & $451\,940$ & 0.89x & 1.23x & 1.47x & 1.47x \\
 \hline
 Random &  $419\,736$ & 1.48x & 2.38x & 3.07x & 3.19x \\
  \hline
\end{tabular}
\end{center}


\bibliography{references}

\appendix
\section{Mathematical model}
\label{math-section}

\subsection{Sequential execution}

Let us estimate how much time is spent on executing $T$ operations sequentially on a binary search tree. Suppose our binary search tree is \emph{external}, i.e., data is contained only in leaves, while internal nodes maintain only routing information. Suppose tree contains $N$ keys and the tree is balanced, therefore the tree height is $O(\log N)$. We suppose \emph{uniform workload}: all keys from the tree are accessed uniformly at random.

\begin{figure}[H]
  \centering
  \caption{Upper levels of the tree are cached, while lower levels reside in RAM}
  \includegraphics[width=\linewidth]{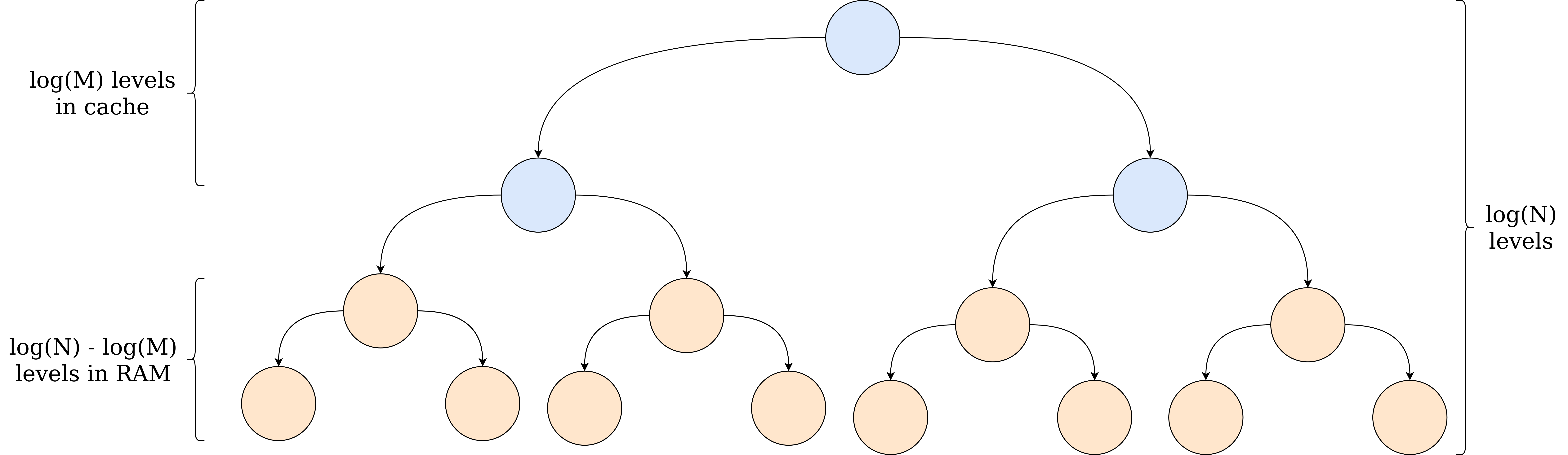}
  \label{tree-ram-cache-pic}
\end{figure}

Suppose the cache size is $M = O(N^{1 - \varepsilon})$, therefore, approximately upper $\log M$ levels of the tree are cached, while $\log N - \log M$ lower levels of the tree are not (Fig.~\ref{tree-ram-cache-pic}).

Each operation first loads $\log M$ nodes from the cache, spending $1$ time unit per each cache fetch. After that, it loads $\log N - \log M$ nodes from the RAM, spending $R$ time units per RAM fetch.

Thus, the sequential execution will take \\ $T \cdot \left( \log M + R \cdot \left( \log N - \log M \right) \right)$ time units to finish, where $T$ is the number of operations.
 
\subsection{Concurrent execution}

Suppose we have $P$ concurrent processes $\{t_i\}_{i = 1}^P$ executing operations concurrently, while each process has its own cache of size larger than $\log N$.

\begin{figure}[H]
  \centering
  \caption{Each successful try of an operation causes unsuccessful tries of $p - 1$ operations}
  \includegraphics[width=0.5\linewidth]{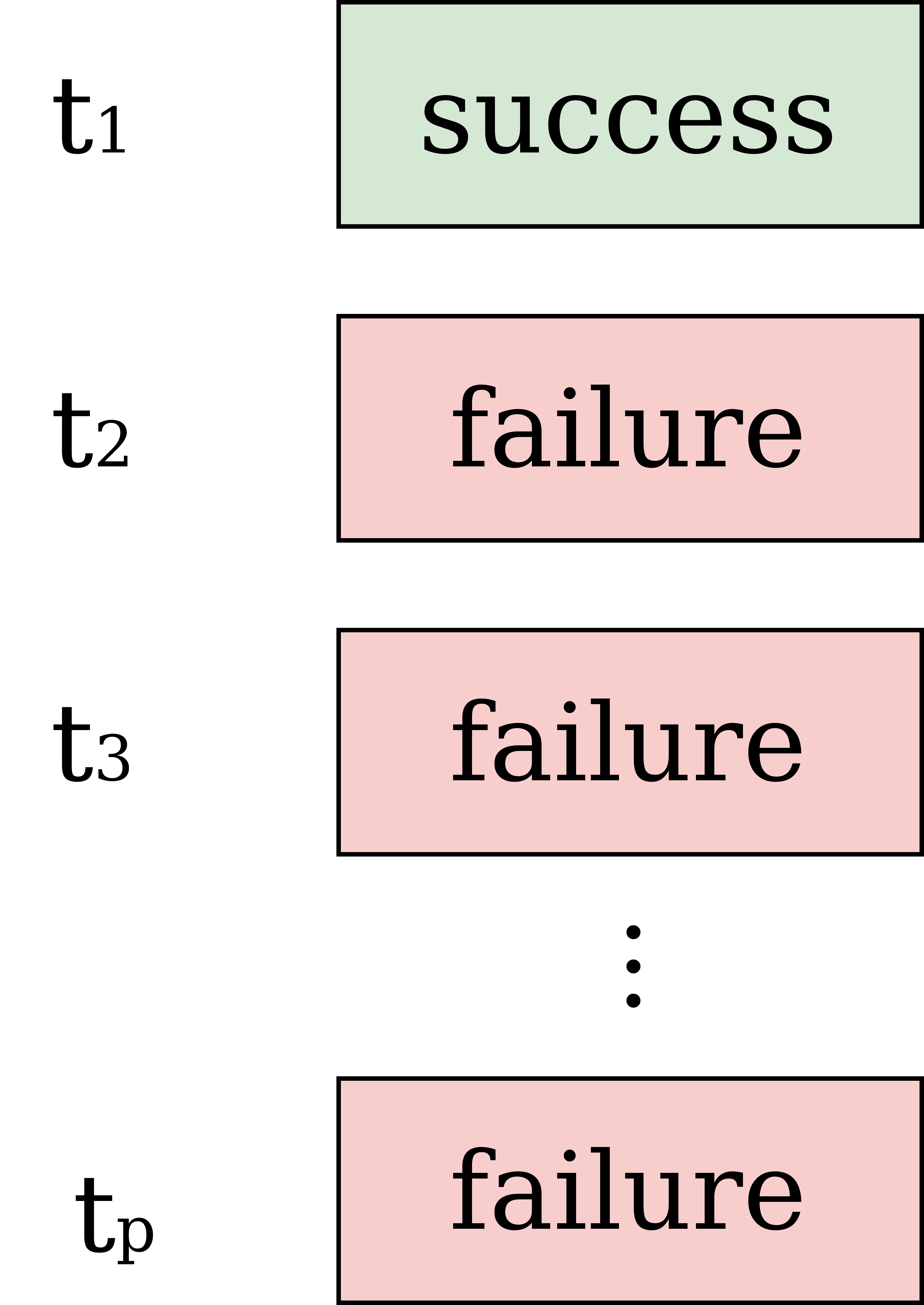}
  \label{p-failures-1-succes-pic}
\end{figure}

In our model we assume that each successful try of a modifying operation causes $p - 1$ unsuccessful tries of modifying operations on other processes (Fig.~\ref{p-failures-1-succes-pic}). 

\begin{figure}[H]
  \centering
  \caption{Nearly each successful modifying operation consists of $P$ retries: $P - 1$ unsuccessful and one successful}
  \includegraphics[width=\linewidth]{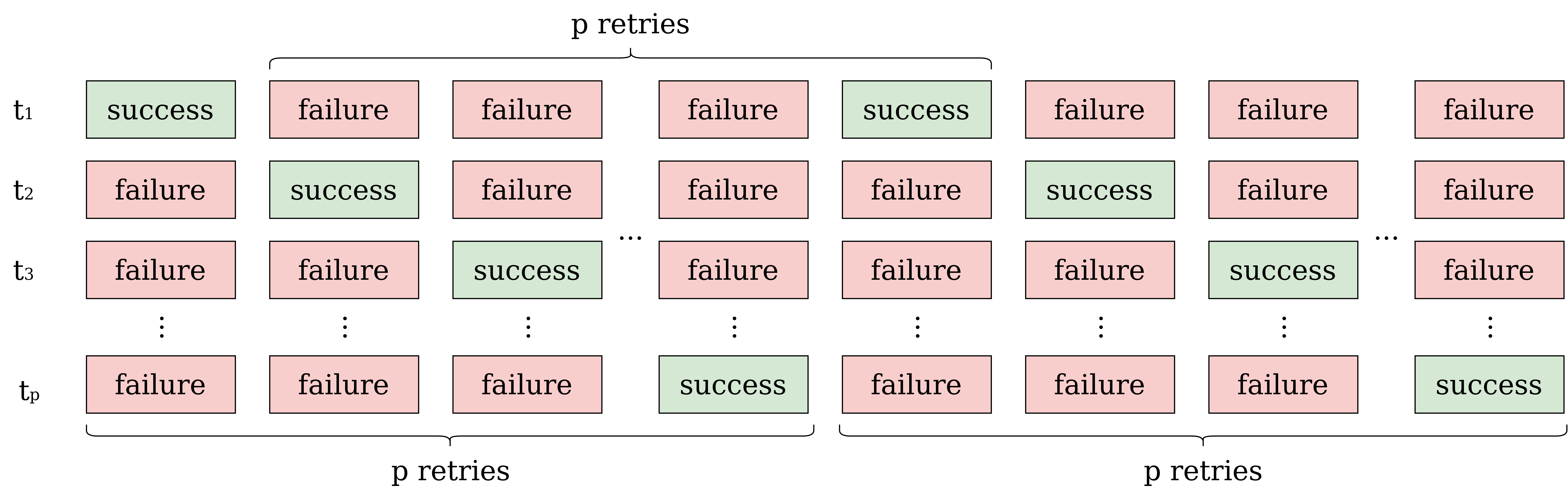}
  \label{retries-diagram-pic}
\end{figure}

We also assume that operation completion events are distributed among processes in a round-robin pattern: first process $t_1$ executes its successful try of the operation, then process $t_2$ executes its successful try of the operation, and so on. Finally, $t_P$ manages to complete its operation, the next process to get its successful try is yet again $t_1$ (Fig.~\ref{retries-diagram-pic}).

As follows from the diagram, almost each successful try of an operation is preceded by $P - 1$ unsuccessful retries (except for $P - 1$ first successful operation, which are preceded by the lower number of unsuccessful retries).

Let us estimate, how long the first retry takes to execute. We must load $\log N$ nodes, none of which might be cached. Thus, we spend $R \cdot \log N$ time units on the first retry.

\begin{figure}[H]
  \centering
  \caption{The number of modified nodes on the path to the requested node}
  \includegraphics[width=\linewidth]{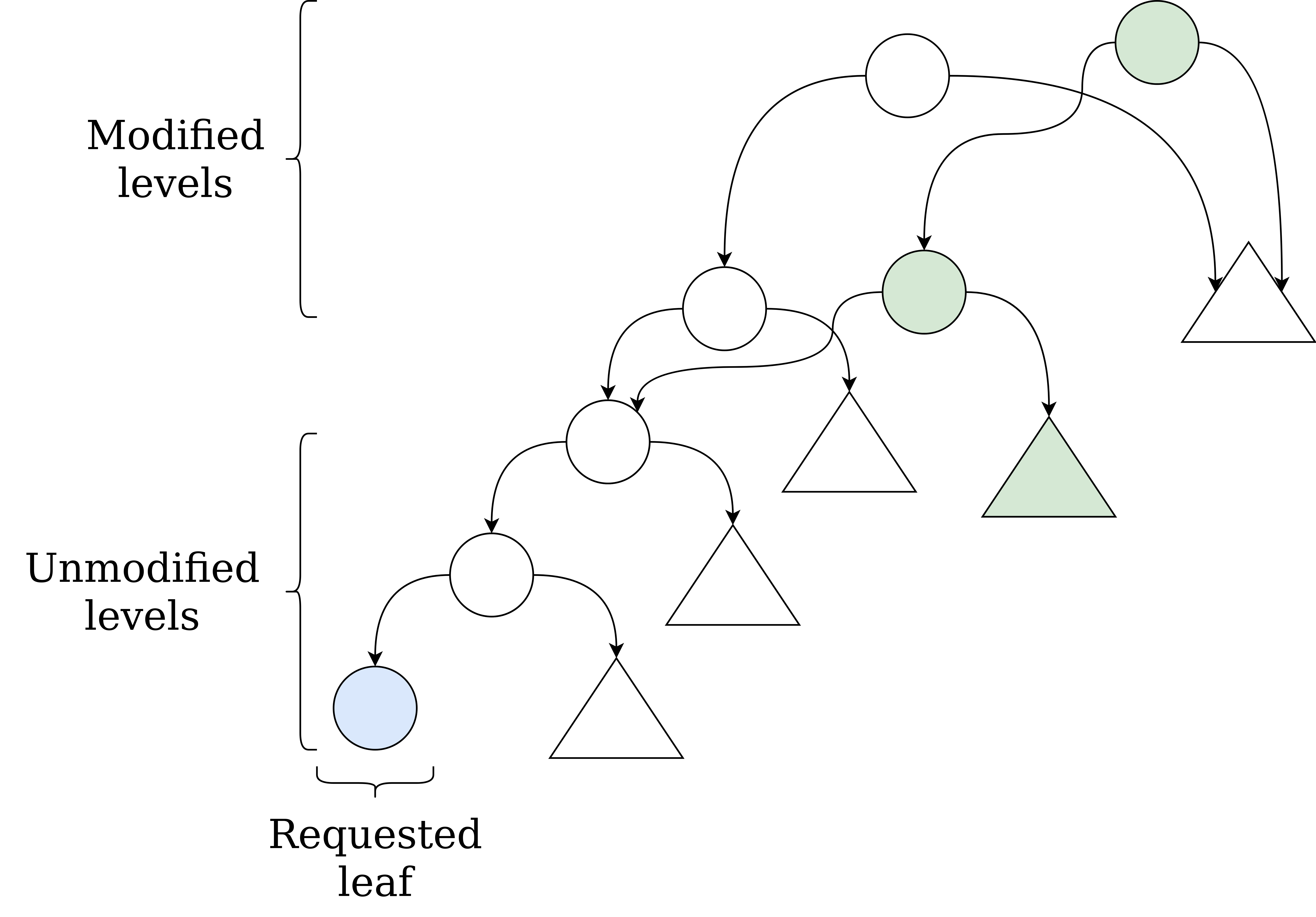}
  \label{modified-levels-pic}
\end{figure}

Let us estimate now how much time we spend on subsequent retries. We begin with estimating, how many nodes on the path to the requested leaf have been modified (Fig.~\ref{modified-levels-pic}). 

Consider the successful modifying operation $op$, that led to a latest failure of our \texttt{CAS} and made us retry our operation the last time. Remember, that arguments of operations are chosen uniformly at random, therefore:

\begin{itemize}
    \item There is $\frac{1}{2}$ probability that $op$ modified some leaf from \texttt{Root->Right} subtree, thus, the number of modified nodes on our path is $1$;
    \item Similarly, there is $\frac{1}{4}$ probability that the number of modified nodes on our path is $2$;
    
    $\ldots$
    
    \item Similarly, there is $\frac{1}{2^k}$ probability that the number of modified nodes on our path is $k$.
\end{itemize}

Thus, we can calculate the expected number of modified nodes on our path $\sum\limits_{k = 1}^{\log N} \frac{k}{2^k} \leq \sum\limits_{k = 1}^{\infty} \frac{k}{2^k} = 2$. Thus, the expected number of modified nodes on our path is not greater than $2$.

Modified nodes were created by another process, thus they do not exist in our process cache. Therefore, they should be loaded out-of-cache, while all the remaining nodes reside in the local cache and can be loaded directly from it. Therefore, we spend $2 \cdot R$ time on average to load all the necessary nodes. In addition, we spend $\log N - 2$ time on average to load all the necessary nodes from the the local cache. Therefore, we spend $2 \cdot R + \log N - 2$ time to fetch all the nodes required for a last operation retry.

An operation execution consists of the first retry, executed in $R \cdot \log N$ and $P - 1$ subsequent retries executed in $(P - 1) \cdot (2 \cdot R + \log N - 2)$. Thus, a single operation is executed in $R \cdot \log N + (P - 1) \cdot (2 \cdot R + \log N - 2)$.

Therefore, we execute $T$ operations in \\ $\frac{T \cdot R \cdot \log N + T \cdot (P - 1) \cdot (2 \cdot R + \log N - 2)}{P}$ time, since we execute these operations in parallel on $P$ processes.

To measure the speedup we simply divide the sequential execution time by parallel execution time: \\
$\frac{T \cdot \left( \log M + R \cdot \left( \log N - \log M \right) \right)}{\frac{T \cdot R \cdot \log N + T \cdot (P - 1) \cdot (2 \cdot R + \log N - 2)}{P}} = P \cdot \frac{\log M + R \cdot \left( \log N - \log M \right)}{R \cdot \log N + (P - 1) \cdot (2 \cdot R + \log N - 2)}$. This gives us $\Omega(\log N)$ speedup when $P = \Omega(min(R, \log N))$ and $R = \Omega(\log N)$.
 
\section{Experiments on other processors}
\label{sec:aux:exp}
We did the same experiments on Intel Xeon Platinum 8160 with 24 cores and AMD EPYC 7662 with 64 cores.



\begin{center}
\begin{tabular}{|c|c|c|c|c|c|} 
 \hline
 Workload & Seq Treap & UC 1p & UC 6p & UC 12p & UC 23p \\
 \hline
 Batch & $638\,600$ & 0.93x & 1.31x & 1.37x & 1.08x \\
 \hline
 Random &  $487\,161$ & 1.24x & 3.23x & 3.55x & 2.8x \\
  \hline
\end{tabular}
\captionof{table}{Results for Intel Xeon Platinum 8160.}
\end{center}



\begin{center}
\begin{tabular}{|c|c|c|c|c|c|c|} 
 \hline
 Workload & Seq Treap & UC 1p & UC 8p & UC 16p & UC 32p & UC 63p \\
 \hline
 Batch & $459\,580$ & 0.96x & 1.7x & 1.91x & 1.55x & 1.02x \\
 \hline
 Random &  $396\,898$ & 1.36x & 3.63x & 2.41x & 2.81x & 2.3x \\
  \hline
\end{tabular}
\captionof{table}{Results for AMD EPYC 7662.}
\end{center}

Unfortunately, one can see that the results are not so impressive when the number of processes is large enough. We suggest that the bottleneck for our benchmarks occurs in Java memory allocator.

\end{document}